\documentclass[aps,pre,twocolumn,floatfix,showpacs,amsmath,amssymb]{revtex4}
\usepackage{epsf}
\usepackage{subfigure}
\usepackage{amsmath}
\usepackage{amssymb}
\usepackage{bm}
\usepackage{graphicx}
\usepackage{epstopdf}
\newcommand{\mean}[1]{\left \langle #1 \right \rangle}

\begin{document}

\newcommand{\Q}{{\widetilde Q}}

\newcommand{\be}{\begin{equation}}
\newcommand{\ee}{\end{equation}}
\newcommand{\bea}{\begin{eqnarray}}
\newcommand{\eea}{\end{eqnarray}}
\newcommand{\cum}[1]{\left \langle \left \langle #1 \right \rangle 
\right \rangle}

\renewcommand{\d}{\text{d}}
\newcommand{\e}{\text{e}}
\renewcommand{\i}{\text{i}}
\newcommand{\rint}{\int\displaylimits}
\renewcommand{\Re}{\mathop{\text{Re}}\nolimits}
\newcommand{\tr}{\mathop{\text{tr}}\nolimits}
\newcommand{\Tr}{\mathop{\text{Tr}}\nolimits}
\newcommand{\ket}[1]{|{#1}\rangle}
\newcommand{\bra}[1]{\langle{#1}|}
\newcommand{\bras}[2]{{}_{#2}\hspace*{-0.2mm}\langle{#1}|}
\newcommand{\bracket}[2]{\langle#1|#2\rangle}
\newcommand{\Det}{\mathop{\text{Det}}\nolimits}
\newcommand{\pv}{\mathop{\text{P}}\nolimits}
\newcommand{\erf}{\mathop{\text{erf}}\nolimits}
\newcommand{\erfc}{\mathop{\text{erfc}}\nolimits}
\newcommand{\erfi}{\mathop{\text{erfi}}\nolimits}
\newcommand{\sinc}{\mathop{\text{sinc}}\nolimits}
\newcommand{\paren}[1]{\left( #1 \right)}
\renewcommand{\labelenumi}{(\roman{enumi})}
\renewcommand{\labelitemii}{$\triangleright$}

\newcommand{\cmt}[1]{\textbf{[#1]}}

\title{\bf Firing Rate of Noisy Integrate-and-fire Neurons with Synaptic Current Dynamics}

\author{David Andrieux}
 \altaffiliation[Also at the ]{Center for Nonlinear Phenomena and Complex Systems, Universit\'e Libre de Bruxelles, CP 231, B-1050 Belgium.}
\affiliation{Department of Neurobiology and Kavli Institute for Neuroscience, Yale University School of Medicine, New Haven, CT 06510, USA
}

\author{Takaaki Monnai}
\affiliation{Department of Applied Physics, Waseda University, 3-4-1 Okubo, Shinjuku-ku, Tokyo 169-8555, Japan
}

\begin{abstract}
We derive analytical formulae for the firing rate of integrate-and-fire neurons endowed with realistic synaptic dynamics. In particular we include the possibility of multiple synaptic inputs as well as the effect of an absolute refractory period into the description. 
\end{abstract}


\maketitle

\section{Introduction}

In vivo neurons in cortical and other neural circuits experience a large background of synaptic inputs, acting as a source of noise. Noisy inputs have an important impact on the dynamics of neurons, making neural responses highly variable and affecting many of their response characteristics \cite{RWRB97,G00,BCFA01,LSL02,CAR02}.  A fundamental problem is thus to determine the output statistics of the neuronal activity given an input noise statistics. 
In addition, the knowledge of the neuronal firing properties can be used to explore large-scale networks using a mean-field approach. In this framework the stationary states of populations of interacting neurons are self-consistently obtained from their firing responses \cite{AB97,B00, BW01}. This allows the efficient exploration of the parameters space and the characterization of the various regime of functioning of these networks. For these reasons it appears crucial to have an accurate estimation of the input-ouput relationship of neurons, especially, in presence of realistic synaptic currents.

In this direction we investigate the firing rates of integrate-and-fire (IF) neurons. The firing frequency of neurons with instantaneous synaptic inputs was first obtained in Ref. \cite{CR71}.
The effect of synaptic dynamics has been studied under various assumptions \cite{BS98,FB02,RP04,RP05}. Here we derive exact formulae that include additional important features. First we take into account the presence of a finite refractory time, which leads to current correlations between spikes affecting the firing rate of the neuron. Indeed, neurons at their reset potential will
evolve with a synaptic current that is still correlated with the {\it positive} going current that made them cross the threshold. 
Furthermore, we also consider the case of multiple inputs from different synaptic receptors, as it occurs in the vast majority of cortical circuits. 

The obtention of the firing rate can be recast into the form of a mean first passage time (MFPT) calculation. Since the synaptic dynamics here plays a central role we have to consider a multi-dimensional Fokker-Planck equation. The nonequilibrium distribution of the synaptic currents at the reset potential implies that the proper phase space boundary conditions must be found self-consistently from the neuronal and synaptic dynamics. To cope with these issues we extend the recently developed tools by Doering and coworkers \cite{DHL87, HDL89,K95}. In this approach the relevant parameter is the ratio between the synaptic current and membrane potential time constants.

Similar considerations also appear in various areas of science when studying the escape rate from a metastable state.
The present results are of special interest in the context of stochastic resonance or stochastic activation \cite{GHJM98, R02}, where considerable attention has been paid to the coexistence of several colored noises with nonequilibrium distributions.

\section{IF neurons and synaptic dynamics}
The sub-threshold neuronal dynamics of integrate-and-fire neurons is described by the depolarization $V(t)$ of the soma, which evolves according to an evolution equation of the form \cite{T88} 
\begin{eqnarray}
\tau_m \frac{dV}{dt} = f(V) +  g^{-1} I(t) 
\end{eqnarray}
with $\tau_m$ the membrane time constant. The function $f(V)$ governs the dynamics of the membrane voltage when no synaptic currents are present. For $f=0$ we have a perfect integrate-and-fire neuron; for $f(V) = -V$ we have a leaky integrate and fire neuron. $I$ denotes the synaptic current and $g$ its associated input conductance. 
The synaptic current evolves according to 
\begin{eqnarray}
\tau_s \frac{dI}{dt} = -I + \mu + \sqrt{D} \xi(t) \, ,
\label{synaptic.current}
\end{eqnarray}
where $\xi$ denotes a Gaussian white noise with zero mean and unit variance:
$\mean{\xi(t)} =0 \quad {\rm and} \quad \mean{\xi(t)\xi(t')} = \delta(t-t')$.
The synaptic current is thus exponentially correlated in time with a correlation time $\tau_s$:
\begin{eqnarray}
\lim_{t\rightarrow \infty} \mean{\bar{I}(t)\bar{I}(t+\tau)} = \frac{D}{2 \tau_s} \exp(-|\tau|/\tau_s) \, ,
\end{eqnarray}
in terms of $\bar{I}(t) = I(t) -\mu$, where $\mu$ is the average current, and the noise intensity $D$. This form of the synaptic dynamics arises when the neuron receives a large number of inputs during its characteristic time, as it is typically the case in vivo in the cortex \cite{T88}. 
In the limit of instantaneous synaptic current, $\tau_s\rightarrow 0$, or for times much larger than the correlation time $\tau_s$, the dynamics simplifies to $\tau_m dV/dt = f(V) +  \mu/g + \sqrt{D}/g \, \xi(t) $. 
Note that the adjunction of the synaptic dynamics \eqref{synaptic.current} renders the voltage dynamics by itself non-Markovian. 

The neuron fires an action potential when the voltage reaches the threshold potential $V_T$. After an absolute refractory period of duration $\tau_r$ it is reset at the value $V_R < V_T$.
During the refractory period no further firing can occur. 
By contrast, even when the neuron is in the refractory state, the synaptic current continues to evolve under Eq. \eqref{synaptic.current}, leading to current correlations of order $\exp \paren{-\tau_r/\tau_s}$ between spikes. This source of correlations must be taken into account in order to obtain a self-consistent input-output relationship.

In the following we will work in the reduced variables
\begin{eqnarray}
z= \epsilon (I-\mu)/\sigma  \quad {\rm and} \quad v=gV \, ,
\end{eqnarray}
where we introduced the parameters
\begin{eqnarray}
\epsilon = \sqrt{\tau_s/\tau_m}  \quad {\rm and} \quad \sigma^2 = D/2\tau_m \, .
\end{eqnarray}
Using the adimensional time $t^{{\rm new}} = t/\tau_m$
we obtain the system of equations
\begin{subequations} \label{evol.reduced}
\begin{eqnarray}
\label{evol.v}
\dot{v} &=& -u'(v) +  \frac{\sigma}{\epsilon}z \\
\label{evol.z}
\dot{z} &=& -\frac{z}{\epsilon^2} + \frac{\sqrt{2}}{\epsilon} \xi(t) \, .
\end{eqnarray}
\end{subequations}
The potential $u(v)$ is such that 
$u'(v)= -gf(v/g) - \mu$,
where the prime denotes a derivative with respect to $v$.
The threshold and reset potentials become
\begin{eqnarray}
\theta \equiv g V_T \quad {\rm and}  \quad \eta \equiv g V_R 
\end{eqnarray}
in these new variables. 
Henceforth we will also refer to the variables $v$ and $z$ as the voltage and the synaptic current, respectively.

\section{Stationary firing rate}

The process \eqref{evol.reduced} obeys the Fokker-Planck equation \cite{R84}
\begin{eqnarray}
\partial_t P  = \Big[\epsilon^{-2} \partial_z(z+\partial_z) - \epsilon^{-1} \sigma z \partial_v + \partial_v u' \Big] P   
\end{eqnarray}
for the joint probability distribution $P(v,z,t)$. We will treat $\epsilon$ as a small parameter, considering the situation where the synaptic time scale is smaller than the membrane time constant. This is for example the case for AMPA receptors, which constitute the main fast excitatory inputs in the brain and have a time constant $\tau_{{\rm AMPA}} \sim 2$ ms smaller than the time constant of pyramidal cells, $\tau_m \sim 25$ ms \cite{DMS98}.  Note that the variable $z$ keeps a finite variance irrespective of the value of the parameter $\epsilon$.

Following Doering et al \cite{DHL87} we have a singular perturbation problem for the quantity 
\begin{eqnarray}
Q(v,z) = \int_0^\infty P(v,z,t) dt \, .
\end{eqnarray}
The latter gives the mean time a neuron spends at points $(v,z)$ before crossing the threshold. 
Alternatively, this stationary problem corresponds to the situation where we perform a time average over a long trajectory where the neuron restarts its time evolution at the reset potential after firing and after its absolute refractory period. After some transients the time spent at each point in phase space will reach the stationary value $Q(v,z)$ \cite{BL82}. The mean first passage time before firing then reads
\begin{eqnarray}
\mean{T}_{\tau_r}=  \int_{-\infty}^{+\infty}  dz  \int_{-\infty}^{\theta}  dv \, Q(v,z)  
\label{MFPT}
\end{eqnarray}
while the firing rate of the neuron is given by
\begin{eqnarray}
\Phi = \frac{1}{\tau_r + \mean{T}_{\tau_r}} \, .
\label{FR}
\end{eqnarray}
The factor $\tau_r$ accounts for the lowering of the firing rate due to the time spent in the refractory state. Importantly, we must consider the additional dependance on the refractory time that appears through the MFPT itself.

To obtain the mean first passage time \eqref{MFPT} we first generalize the calculation of Doering et al \cite{DHL87} to the case where the synaptic current at reset potential follows an arbitrary distribution $\mu (z)$. This distribution will have to be determined self-consistently from the neuronal dynamics. The function  $Q(v,z)$ obeys the equation 
\begin{eqnarray}
 \Big[\epsilon^{-2} \hat{L} - \epsilon^{-1} \sigma z \partial_v + \partial_v u' \Big] Q(v,z)  = - \delta(v-\eta) \mu(z) \, , \,
\label{G.eq}
\end{eqnarray}
where we introduced the operator $\hat{L}\equiv \partial_z(z+\partial_z)$.
Furthermore, it must satisfy the half-line absorbing boundary condition \cite{WU45}
\begin{eqnarray}
Q(\theta,z) = 0 \quad {\rm for} \quad  z < \frac{\epsilon u'(\theta)}{\sigma} \, .
\label{BC}
\end{eqnarray}
This condition stems from the observation that the probability current $J_v= \paren{-u'(v)+\sigma z/\epsilon }Q$ must be positive at the threshold potential for firing to occur.
Intuitively this corresponds to the fact that the potential cannot cross the threshold from above.

For further reference we here introduce the eigenfunctions of the operator $\hat{L}$:
\begin{eqnarray}
\hat{L}\rho_n (z)= - n \rho_n(z)\, , \quad \rho_n (z) = \frac{e^{-z^2/2}}{\sqrt{2\pi}} {\rm He}_n (z) \, ,
\end{eqnarray}
where ${\rm He}_n (z)= (-1)^n e^{z^2 /2} (d^n/dz^n) e^{-z^2 /2}$ are the Hermite polynomials \cite{AS70}. The latter form a complete orthogonal basis in the inner product $\int_{-\infty}^{+\infty} {\rm He}_n(z) {\rm He}_m(z)\exp(-z^2 /2) = \sqrt{2\pi} n! \,\delta_{nm}$. Hence we may introduce the quantities
\begin{eqnarray}
c_n = \frac{1}{n!} \int_{-\infty}^{+\infty} \mu (z) {\rm He}_n (z) \, dz 
\end{eqnarray}
characterizing the distribution $\mu$.
Note that $c_0 = \int_{-\infty}^{+\infty} \mu (z) \, dz = 1$ and that $c_1 = \int_{-\infty}^{+\infty} \mu (z) z \, dz$ is the mean of the distribution $\mu$. 

We now insert the ansatz 
\begin{eqnarray}
Q(v,z) = Q_0(v,z) + \epsilon Q_1(v,z) +  \epsilon^2 Q_2(v,z) + \dots 
\label{ansatz}
\label{Q.series}
\end{eqnarray} into Eq. \eqref{G.eq} and collect terms according to powers of $\epsilon$.
The zeroth order corresponds to the white noise limit of Eqs. \eqref{evol.reduced}, i.e., to the limit where synaptic events are instantaneous. We have
\begin{eqnarray}
\hat{L} Q_0 (v,z)= 0
\end{eqnarray}
whose solution is
\begin{eqnarray}
Q_0 (v,z)= r_0 (v)\rho_0(z) \, ,
\end{eqnarray}
with a yet undetermined function $r_0(v)$.  
At the next order of perturbation we find
\begin{eqnarray}
\hat{L} Q_1 (v,z)= \sigma \rho_1(z) \partial_v r_0(v) \, ,
\end{eqnarray}
yielding
\begin{eqnarray}
Q_1 (v,z)= r_1 (v)\rho_0(z)- \sigma \rho_1(z) \partial_v r_0 (v) 
\end{eqnarray}
with the yet unknown function $r_1(v)$. 
The second-order terms give
\begin{eqnarray}
\hat{L}Q_2 (v,z)= - \delta(v-\eta) \sum_{n=0}^\infty c_n \rho_n(z) +\rho_1(z)\sigma  \partial_v r_1(v)  \nonumber \\
- \rho_2(z)\sigma^2 \partial_v r_0(v) -\rho_0(z) [\partial_v(u' r_0) +  \sigma^2 \partial_v r_0(v)] \, , \, \, 
\end{eqnarray}
where we used that $z\rho_n (z) = \rho_{n+1} (z)+ \rho_{n-1}(z)$.
The operator $\hat{L}$ is invertible only in the subspace of functions spanned by $\rho_n$, $n \geq 1$. Hence  the term proportional to $\rho_0$ must vanish, leading to the following equation for $r_0(v)$:
\begin{eqnarray}
\sigma^2 \partial^2_v r_0 + \partial_v(u' r_0) =  -  c_0 \delta(v-\eta)
\end{eqnarray}
with $c_0=1$. Similarly, the integrability condition for $Q_3$ leads to the following equation for $r_1(v)$:
\begin{eqnarray}
\sigma^2 \partial^2_v r_1 + \partial_v(u' r_1) =  - c_1 \sigma \partial_v \delta(v-\eta) \, .
\end{eqnarray}
The solutions of these equations that are integrable read
\begin{eqnarray}
r_0 (v) &=&  \sigma^{-2} e^{-u(v)/\sigma^2} \int_\theta^v dv' \, e^{u(v')/\sigma^2} \Theta(v'-\eta) \nonumber \\ &+& A \sigma^{-2} e^{-u(v)/\sigma^2} 
\label{r0}
\end{eqnarray}
and
\begin{eqnarray}
r_1 (v) &=& B \sigma^{-1} e^{[u(\theta)-u(v)]/\sigma^2} \qquad \qquad \nonumber \\ &+& c_1 \sigma^{-1} e^{[u(\eta)-u(v)]/\sigma^2} \Theta(\eta-v) \, , 
\label{r1}
\end{eqnarray}
where $\Theta$ is the Heaviside function. Remarkably the first-order correction to the MFPT, and hence to the firing rate, only depends on $c_1$, that is on the mean of the current distribution $\mu(z)$. 
Note that the MFPT can be expressed as $\mean{T} = \int_{-\infty}^\theta [r_0(v) + \epsilon r_1(v) + \cdots] dv$ as the functions $r_n(v)$ are the reduced densities, $r_n(v)=\int_{-\infty}^{+\infty} Q_n (v,z) dz$.

We now have to determine the factor $c_1$ and the constants $A$ and $B$ self-consistently. In this regard, we first consider the phase space boundary condition \eqref{BC}.
The latter was solved in Ref. \cite{HDL89} and can be summarized by the condition
\begin{eqnarray}
r(\theta) = \epsilon \sigma \alpha r'(\theta)
\end{eqnarray}
where $\alpha \equiv - \zeta (1/2)\simeq 1.46\dots$ with $\zeta$ the Riemann zeta function. 
The crucial point to observe is that this boundary condition remains valid in presence of an arbitrary current distribution $\mu$ since, as shown by Eq. \eqref{r1}, the presence of a non-vanishing mean current, $c_1 \not= 0$, only affects the 
residence time in the region $v< \eta$, leaving the region $\eta < v < \theta$ unchanged.
Hence the constants $A$ and $B$ take the values $A=0$ and $B=\alpha$. The factor $\alpha$  is referred to as the Milne extrapolation length as it characterizes the non-vanishing value of the probability distribution at the threshold potential.

We now have to evaluate the value of the mean current $c_1$ at the reset potential.
It must be determined self-consistently as the synaptic current remains correlated between spikes due to its finite time constant. Accordingly, the probability distributions at the threshold and reset potentials are related as follows:
\begin{eqnarray}
Q (\eta,z) = \int_{-\infty}^{+\infty}   G(z,z',\tau_r) Q (\theta,z') \, dz' 
\label{match}
\end{eqnarray}
where $G(z,z',t)$ is the Green function of the synaptic current, which is given by a normal distribution for $z$ of mean $z' \exp(-t/\epsilon^2)$ and variance $1-\exp(-2t/\epsilon^2)$.
Consequently we have that 
\begin{eqnarray}
c_1 = \exp(-\tau_r/\tau_s) \int_{-\infty}^{+\infty} z' Q (\theta,z') dz' \, ,
\end{eqnarray}  
where we used Eq. \eqref{match} with $\int_{-\infty}^{+\infty} z G(z,z',\tau_r)dz = z'\exp(-\tau_r/\tau_s)$. 
The mean reset current is thus related to the mean threshold current. Again, we can verify that, at first-order in $\epsilon$, the expression for the probability flux given in Ref. \cite{HDL89} remains unchanged in the present situation, as already suggested by Eq. \eqref{r1}. Hence we find 
\begin{eqnarray}
c_1 = \alpha \exp(-\tau_r/\tau_s)
\end{eqnarray}
at first-order in $\epsilon$. When the refractory time is large compared to the synaptic time constant the current will relax to its stationary distribution $\mu (z) =  e^{-z^2/2}/\sqrt{2\pi}$ so that $c_n =0$ for $n\geq 1$. When the refractory time vanishes correlations are maximal since the current distribution $\mu$ at the reset potential exactly matches the current distribution at the threshold, resulting in a contribution $c_1 = \alpha$. The same method can be applied to obtain the full distribution of the synaptic current. 

Combining these results we obtain the firing rate \eqref{FR} of IF neurons. An important case is the leaky integrate-and fire neuron for which the mean first passage time takes the compact form
\begin{eqnarray}
\mean{T} = \tau_m \sqrt{\pi} \int_{\eta^*}^{\theta^*}  e^{w^2}[{\rm erf}(w)+1] dw
\label{MFPTIF}
\end{eqnarray}
where $\eta^*=(V_R-\mu/g)/\sqrt{2}\sigma + \epsilon \alpha \exp(-\tau_r/\tau_s)/\sqrt{2}$, $\theta^* = (V_T-\mu/g)/\sqrt{2}\sigma+ \epsilon \alpha/\sqrt{2}$, and erf is the error function \cite{AS70}. 
This formula interpolates between the two-limiting cases of zero and infinite refraction time previously considered in Refs \cite{FB02} and \cite{BS98}, respectively. Similar expressions can be obtained for nonlinear IF neurons as well.

\begin{figure}[t]
\centerline{\scalebox{0.35}{\includegraphics{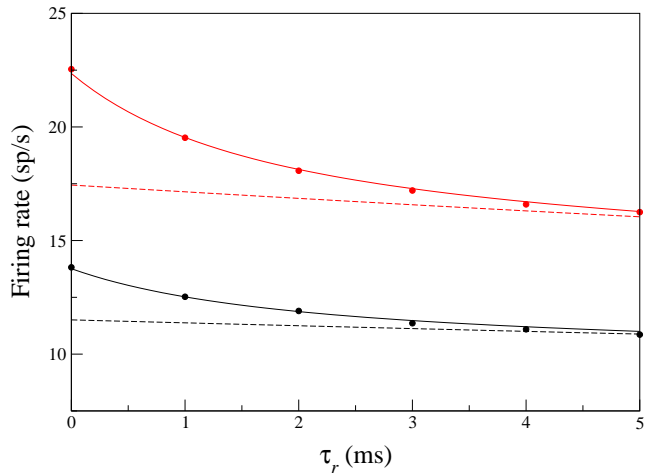}}}
\caption{Firing rate of leaky integrate-and-fire neurons as a function of the absolute refractory period. The dashed lines represent the approximation of stationary synaptic current (see text). Dots depict the result of Monte-Carlo simulations. The reset and threshold potentials take the values $V_T = 15$ mV and $V_R = 20$ mV, respectively (the resting potential is $0$ mV). The membrane time constant $\tau_m = 25$ ms and the synaptic time constant $\tau_s = 2$ ms so that $\epsilon \approx 0.283$. The mean potential $\mu/g = 15.5$ mV and its standard deviation $\sigma =5$ mV for the lower curve (black), and $\mu/g = 16.5$ mV and $\sigma =6$ mV for the upper curve (red).}
\label{fig1}
\end{figure}

The firing rate \eqref{FR} as a function of the refractory period is depicted in Fig. \ref{fig1} along with the result of Monte-Carlo simulations. The firing rate is compared to the situation where the neuron is unresponsive during the refractory period but the corresponding current correlations are neglected [i.e., formula \eqref{FR} but with $\mean{T}_\infty$]. The corrections due to the effect of the refractory state on the MFPT are significant, even at moderate firing rates. 

\begin{figure}[t]
\centerline{\scalebox{0.35}{\includegraphics{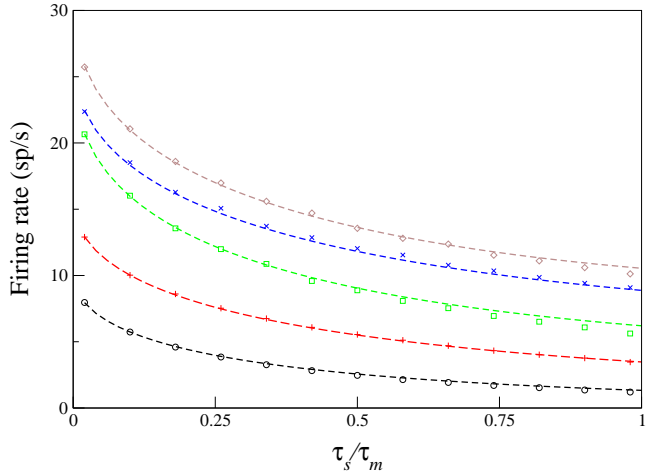}}}
\caption{Firing rate of leaky integrate-and-fire neurons as a function of the ratio $\tau_s/\tau_m$. Monte-Carlo simulations (symbols) are compared to formula \eqref{FR} with the effective threshold \eqref{effthresh} and $\tau_r =0$ (dashed lines). Different curves correspond to different means and variances of the synaptic input.}
\label{fig2}
\end{figure}

To investigate the validity of the expansion (\ref{ansatz}), we numerically obtained the firing rate as a function of the ratio $\tau_s/\tau_m$, which is plotted in Fig. \ref{fig2}. The agreement with the first-order formula \eqref{MFPTIF} holds for the region $\tau_s/\tau_m < 0.1$. For ratios $\tau_s/\tau_m$ up to $1$, we found that the firing rate could be well fitted by considering the effective threshold
\begin{eqnarray}
\theta^* = \frac{(V_R-\mu/g + 0.1375 \epsilon^2)}{\sqrt{2}\sigma} + \epsilon \frac{\alpha}{\sqrt{2}} - 0.225 \epsilon^2 \, ,
\label{effthresh}
\end{eqnarray}
as seen in Fig. \ref{fig2}. The effective reset potential $\eta^*$ appears unaffected by the second order corrections, as observed in Ref. \cite{BS98}. In particular, the dependance on the refractory period remains identical.  

\section{Multiple synaptic inputs}

We further consider the case where several synaptic inputs are present, i.e., 
\begin{eqnarray}
I = I_1 + I_2
\end{eqnarray}
where the currents $I_k$ obey evolution equations of the form \eqref{synaptic.current} albeit with different time constants $\tau_{s_k}$, means $\mu_k$, and variances $\sigma^2_k$. We here assume that the synaptic time constants satisfy $\tau_{s_1}/\tau_{s_2} = c^2$ with $c$ of order unity, $c = O(1)$, so that $\tau_{s_k}$ are of order $\epsilon$: $\sqrt{\tau_{s_1}/\tau_m} = c\sqrt{\tau_{s_2}/\tau_m}=  \epsilon$. The calculations of the previous section can be extended to this case by considering the three-dimensional phase space Fokker-Planck equation. 
Inserting the eigenfunctions $\rho_n (z_1) \rho_m (z_2)$ into the expansion in powers of $\epsilon$, we obtain that the boundary condition now reads $\epsilon \sigma \lambda r' (0) = r(0)$, where $\sigma^2 = \sigma_1^2 + \sigma_2^2$.
We could not obtain the Milne extrapolation length analytically, which was determined numerically to be $\lambda \approx 1.4$. It replaces the factor $\alpha=-\zeta(1/2)\simeq1.46 \dots$ in the present situation. 

The white noise limit is obtained from Eq. \eqref{r0} with $C=0$ and $\sigma^2=\sigma_1^2+\sigma_2^2$, and where the potential $u$ incorporates both input currents, $\mu=\mu_1 + \mu_2$. The first-order correction reads
\begin{eqnarray}
r_1 (v) = \lambda e^{[u(0)-u(v)]/\sigma^2} \, ,
\end{eqnarray}
with the corresponding correction to the MFPT given by $\epsilon\int_{-\infty}^{+\infty} r_1(v) dv$ (for simplicity we did not include the refractory state in this analysis). Remarkably, this first-order correction only depends on the total noise strength $\sigma^2$ and not on the ratio $c^2$ of the synaptic time constants. Accordingly, as regards the firing rate, we may treat different synaptic currents as an effective single input with parameters $\mu=\mu_1 + \mu_2$ and $\sigma^2=\sigma_1^2+\sigma_2^2$, even when they present different time courses. This effect is valid at first-order in the parameter $\epsilon$; the case of very long synaptic time constants compared to the membrane integration time was studied in Ref. \cite{RP04}. The generalization to $n>2$ synaptic inputs is straightforward and does not change this result.

\section{Conclusions}
\label{Conclusions}

We have obtained analytical expressions for the stationary firing rate of integrate-and-fire neurons endowed with realistic synaptic dynamics.
Precisely we have included two salient features into the description: the presence of a finite refraction time on the one hand, and the presence of multiple synaptic inputs on the other hand.

The finiteness of the refractory period of the neuron provides a new source of correlations interacting with the synaptic dynamics. 
Since the synaptic dynamics has a finite relaxation time, neurons at the reset potential may still be correlated with the positive current that made them cross the firing threshold, resulting in an increased firing rate. This mechanism is important for fast excitatory and inhibitory synapses such as AMPA and ${\rm GABA}_A$ synapses as their time constants is in the range of $2$ to $5$ ms \cite{DMS98}, comparable to the absolute refractory period of cortical neurons, $\tau_r \sim 1-5$ ms.

The presence of multiple synaptic inputs has been considered as well. In this case the Milne extrapolation length must be determined numerically to complete the theoretical analysis. Importantly the firing rate only depends on the variances of the synaptic inputs and not on their relative time constants. This allows us to consider the combined effect of excitatory and inhibitory synaptic inputs from, say, AMPA and ${\rm GABA}_A$ receptors.

These analytical results are crucial to assess the effect of noise on neuronal dynamics. For instance they provide accurate expressions that can be used in the mean-field exploration of large-scale neuronal networks \cite{AB97,B00, BW01}. In particular, the network collective properties, such as the stability of the low activity spontaneous state or of the persistent "memory" states, will depend on the transfer function. More generally, the study of nonequilibrium, multiple colored noises appears in a wide range of situations of concern, from chemical networks to biology.

\section{Acknowledgments}

D.~Andrieux thanks Professor X-J~Wang for support and encouragement in this research and acknowledges financial support from the F.R.S.-FNRS Belgium. T.~Monnai acknowleges financial support from a Waseda University grant for special projects.

\end{document}